\documentstyle[aps,preprint]{revtex}
\begin{document}
\draft
\title{E. Noether's Discovery of the Deep Connection Between
Symmetries and
                   Conservation Laws\footnote{To be published in
the Proceedings of a
Symposium on the Heritage of Emmy Noether which was held in 
Bar-Ilan University, Isreal Dec. 2-4,
1996.}}

\preprint{UCLA/98/TEP/20; hep-th/9807044}

\author{Nina Byers} \address{Physics
Department, UCLA, Los Angeles, CA 90024} \date{July 16, 1998}
 \maketitle

\begin{abstract} 
Emmy Noether proved two deep theorems, and their converses, on
 the connection between symmetries and conservation laws. Because these
 theorems are not in the mainstream of her scholarly work, 
  which was the development of modern
 abstract algebra,  it is of some historical interest to examine how
 she came to make these discoveries.  The present paper is an historical account of the
 circumstances in which she  discovered and proved  these theorems which
 physicists refer to collectively as Noether's Theorem.  The work was done 
 soon after  Hilbert's discovery of the variational principle which
 gives the field equations of general relativity.  The failure of local energy
 conservation in the general theory was a problem that concerned people at that
 time, among them David Hilbert, Felix Klein, and Albert Einstein.  Noether's
 theorems solved this problem.  With her characteristically deep insight and
 thorough analysis, in solving that problem she discovered very general theorems
 that have profoundly influenced modern physics. 
 \end{abstract}

\section{INTRODUCTION}

Emmy Noether's paper \lq\lq{\it{Invariante Variationsprobleme}}"\cite{noet},
hereafter referred to as {\it{I.V.}}, profoundly influenced 20th century
physics.  It was presented to the July 16, 1918 meeting of the
{\it{K\"{o}nigliche Gesellschaft der Wissenschaften zu G\"{o}ttingen}} by Felix 
Klein and published in its {\it{Nachrichten}} [Proceedings]. It was presented by 
 Felix Klein  presumably because
Noether was not a member of this Royal Society.
{\footnote{The Royal Society of London, founded in 1662, elected its first
female member in 1945; the Acad\'{e}mie des Sciences, Paris, was founded in
1666 and elected its first female member in 1962.}} One wonders if she was even
present when the paper was read.  The paper proved two theorems and their
converses which revealed the general connection between symmetries and
conservation laws in physics.  They led to a deeper understanding of laws such
as the principles of conservation of energy, angular momentum, etc., and also
were instrumental in the great discoveries of gauge field symmetries of the 20th
century.\cite{byers}

Remarkably these theorems are something of a departure from the main line of Noether's
mathematical research, which was the development of modern abstract algebra.  An examination
of the historical circumstances in which she made these discoveries sheds light on how she came to do this work.
  The work  followed upon David Hilbert's discovery
of the variational principle from which he derived
  the field equations of
general relativity.  
At this time 
 David Hilbert, Felix Klein and others in G\"{o}ttingen were intensely  interested in the recently  completed general theory of relativity. There were
unresolved issues regarding the question of energy conservation in the theory.
Published correspondence \cite{klein} 
between Hilbert and Klein regarding this indicates that   Noether's help was requested by Hilbert to clarify these
issues, and that she
then did the work whose results are published a few years later in {\it{I.V.}}. 
The discussion and proofs of  the two theorems  in {\it{I.V.}} clarified and
resolved the issues regarding energy conservation, as Hilbert acknowledged
in his 1924 {\it{Grundlagen der Physik}} article \cite{hilbert1}. From her Collected Works \cite{publist}, it would
seem that after this and some additional work on
differential invariants in general relativity\cite{paper1}, Noether returned to the main line
of her mathematical research.

Here I will present first a chronology of events leading up to Noether's publication of the
{\it{I.V.}} paper;  then a discussion of some rather difficult issues regarding
energy conservation in general relativity which her work resolved; and finally a
description and discussion of the two theorems in the {\it{I.V.}} paper.

       \section{Chronology of Events Leading to the Discovery}

In 1915, Emmy Noether was invited to join the team of mathematicians assembled
in G\"{o}ttingen by David Hilbert. Hermann Weyl reports \cite{weyl}: \lq\lq To
both Hilbert and Klein, Emmy was  welcome as she was able to help them with 
invariant-theoretic knowledge." She was  thirty-three at that time, having
received a doctorate in mathematics from the University of Erlangen seven years
earlier, and  written the eleven  papers listed in Appendix A. 
The first in the list is
 her thesis which was
done under  the supervision of Paul Gordan. For her thesis 
 she 
calculated  all the 331  invariants of ternary bi-quadratic forms!  Shortly
thereafter she took the abstract approach to algebra following 
 Hilbert's 1888 basis theory paper. She worked unpaid in Erlangen supervising
students and sometimes lecturing for her ailing father. After her father died,
she joined Hilbert and his team in G\"{o}ttingen. \cite{Dick}

In June-July 1915,
shortly after Noether arrived,  Albert Einstein gave  six 
lectures in G\"{o}ttingen on the general  theory of relativity.   At that time the theory  was
not yet finished; he had not yet found the complete field equations. However,
the basic ideas were clear and his audience found them  compelling.  
After giving the lectures, Einstein said \cite{pais}: \lq\lq To my great joy, I completely 
succeeded in convincing Hilbert and Klein." He had been working to generalize
the special theory of relativity to include gravity since 1905. In 1907 he
discovered the importance  of the equality of gravitational and inertial mass
and formulated the equivalence principle, but it took another eight years  to
complete the theory.  Finally in November 1915, having found the complete field
equations, he submitted   the famous paper \cite{einstein} ~ that gives the theory
in its final form. Remarkably, in the same month,  Hilbert
submitted a manuscript \cite{hilbert} ~ in which the same field  equations are
obtained as the solution to a variational problem. Hilbert and
Einstein  had independently found the field equations at about the same
time \cite{pais}.  
In  November 1915, Emmy Noether  wrote to Ernst Fischer:
\lq\lq Hilbert plans to lecture next week about his ideas on Einstein's
differential invariants, and so ..[we] had better be ready" \cite{Dick}.  It would
seem, therefore, that she 
 began to study relativity theory then.  Out of that
study came two papers which Hermann Weyl characterized as giving \lq\lq the
genuine and universal mathematical formulation of two of  the most significant
aspects of general relativity theory: first, the reduction of the problem of
differential invariants to a purely algebraic one by use of normal coordinates;
and second, the identities between the left sides of Euler's equations of a
problem of variation ..."\cite{weyl}.  Regarding the first paper 
referred to by Weyl 
\cite{paper1}, Einstein wrote to Hilbert: \lq\lq Yesterday I  received from Miss
Noether a very interesting paper on invariant forms. I am impressed that one
can comprehend these matters from so general a viewpoint. It would not have
done the old guard at G\"{o}ttingen any harm had they picked up a thing or two
from her. ..."\cite{kimb}. The second paper referred to by Weyl
  is the {\it{I.V.}} paper we will discuss at length.

Hilbert had been interested in the fundamental laws of physics for many years \cite{reid}.
The paper in which he published his derivation of the field equations for general relativity
is entitled {\it{Grundlagen der Physik}} \cite{hilbert}.  It was an early effort to formulate
a unified field theory of gravity, electromagnetism, and matter.  In this it was
unsuccessful, and Hilbert omitted this paper from his collected works \cite{hilbert1}.
However, when specialized to gravity, Hilbert's derivation of the field equations is an
original and important contribution to the general theory.  The Lagrangian he introduced is
known as the Hilbert-Einstein Lagrangian, and his formulation of the theory is in widespread
use today.  Pais writes:\lq\lq Hilbert was not the first to apply the principle to
gravitation.  Lorentz had done so before him.  So had Einstein, a few weeks earlier.  Hilbert
was the first, however, to state this principle correctly."  As regards Hilbert's 1915 paper,
Hermann Weyl \cite{reid} wrote:  \lq\lq Einstein's more sober procedure ...[has] proved the
more fertile.  Hilbert's endeavors must be looked upon as a forerunner of a unified field
theory of gravitation and electromagnetism.  However, there was still much too much
arbitrariness involved in Hilbert's Hamiltonian function; subsequent attempts (by Weyl,
Eddington, Einstein himself, and others) aimed to reduce it.  ...  But the problem of a
unified field theory stands to this day as an unsolved problem."

Though the general theory of relativity was completed in 1915, there remained unresolved
problems.  In particular, the principle of local energy conservation was a vexing issue.  In the general theory, energy
is not conserved locally as it is in classical field theories - Newtonian gravity,
electromagnetism, hydrodynamics, etc..  Energy conservation in the general theory has been
perplexing many people for decades.  In the early days, Hilbert wrote about 
this problem as \lq the
failure of the energy theorem '. In a correspondence with Klein \cite{klein}, 
he asserted that this `failure' is  a characteristic feature of the general
theory, and that instead of `proper energy theorems' one had
`improper energy theorems' in such a theory. This conjecture was clarified,
quantified and proved correct by Emmy Noether.  In the note to Klein 
he reports that had requested that Emmy Noether help 
clarify the matter. In the next section this problem will be
described in
more detail and an explanation given of how Noether clarified, quantified, and
proved Hilbert's assertion. One might say it is a lemma of her Theorem II. 

 Klein was working on this problem of energy conservation, or as he
termed it Hilbert's energy vector, in 1916.   He wrote in this correspondence 
with Hilbert that
\lq\lq Frl.  Noether advised me continually throughout my work, and it was really only
through her that I was led to the material presented in this letter.  When I spoke recently
with Frl.  Noether about my result concerning your energy vector, she told me that she had
derived the same thing from your Note\footnote{ Ref.  \cite{hilbert}} a year ago and written
it up in a manuscript ( which I examined )."  Klein presented a paper 
with his results 
 on Hilbert's energy
vector to the July 19 meeting of the {\it{Gesellschaft}}.   In this paper
\cite{kleinpap}, he
acknowledges helpful contributions from Noether. However, in the annotation
to this paper in his
collected works he makes it clear that his result is a special case of her
`far-reaching' theorem.  He reports that he presented her paper to the {\it{Gesellschaft}}
the following week, and  says that she also proved and generalized some of Hilbert's
ideas.  An English translation of that interesting annotation is given in Appendix B.

In his famous
1924 {\it{Grundlagen der Physik}} paper\footnote{ This is included in his Collected Works
\cite{hilbert1}.   It has the same title as the 1915 paper which was omitted from the
 Collected Works.}, David Hilbert credits  Emmy  Noether  with having
solved the problem of an energy theorem in the general  theory, and refers to 
 {\it{I.V.}}. In the concluding section of her paper, she refers to Hilbert's having said that `the failure of
the energy theorem' is a characteristic feature of the general theory. 
 The last section is entitled
 \lq\lq A HILBERTIAN ASSERTION", and says in part  \cite{basil}:  \lq\lq From the foregoing
...[we] obtain the proof of an assertion of Hilbert concerning the connection between the
failure of proper energy conservation laws and general relativity, and indeed in a general
group-theoretic setting."  It seems to me this remark is a notable understatement.  Her powerful results are
 deep and  general; far more  than simply a  verification of Hilbert's assertion.

In 1919, Noether chose {\it{I.V.}} for her {\it{Habilitation}} thesis and presented it
to the University of G\"{o}ttingen along with the twelve previously published
papers and two additional  manuscripts, one of which \cite{schm} contained a
number of important ideas which had a significant impact on the development of
modern abstract algebra. Years before she had been refused {\it{Habilitation}} 
( the right to lecture under University auspices).  Hermann Weyl records
\cite{weyl} \lq\lq during the war Hilbert tried to push through Emmy Noether's
{\it{Habilitation}} in the Philosophical Faculty in G\"{o}ttingen. He failed
due to the resistance of the philologists and historians.  It is a well-known
anecdote that Hilbert ... [declared] at the faculty meeting; \lq I do not see
that the sex of a candidate is an argument against her admission as
{\it{Privatdocent}}. After all, we are a university not a bathing
establishment.' " It was only after the liberalization of Germany in the
post-WWI period that   {\it{ Habilitation}} was granted, and she was finally
able to lecture officially at the University. She had given lectures in the
preceeding years, but the posted notices  stated that they were offered by Herr
Professor David Hilbert,  given by Frl. E. Noether,  with no tuition required.
These  lectures were becoming famous and drawing the attendance of
mathematicians from all over Europe.  Without {\it{Habilitation}} she was
working unpaid.

It is quite clear from                     the application papers written for
{\it{Habilitation}} that Noether knew the scope and importance of her results
for physics. She wrote about the {\it{I.V.}} paper \cite{Dick} \lq\lq ...  The general
results  contain, as special cases, theorems on first integrals [conservation
laws] ... in mechanics; furthermore, conservation theorems and interdependences
among the field equations in the theory of relativity... " She remarks 
that this work was \lq\lq an outgrowth of my assistance to Klein and Hilbert in
their work on Einstein's  general theory of relativity."
\section{`PROPER' AND  `IMPROPER' CONSERVATION LAWS}

In contemporary terminology the general theory of relativity is a gauge theory.
The symmetry group of the theory, is a gauge group. It is the group of all continuous coordinate 
transformations with continuous derivatives,   often called
the group of general coordinate transformations. 
   It is a Lie group that has a continuously infinite number of independent infinitesimal generators. In Noether's terminology  such a group is an infinite continuous group.
The symmetry group of special relativity,  the Poincar\'{e} group\footnote{
The group of Lorentz transformations and spacetime translations.}, 
 is a Lie subgroup of the group of general coordinate transformations.  It
 has a finite number (7) of independent infinitesimal
generators. Noether refers to such a  group as a finite continuous group.
This distinction between a Lie group with a finite ( or countably infinite ) number of independent infinitesimal generators and an infinite continuous
group is what distinguishes Noether's theorem I and theorem II  in {\it{I.V.}}.
Theorem I applies when one has a finite continuous  group of symmetries,
and theorem II  when there is an infinite continuous group of
symmetries. Field theories with a finite continuous symmetry group have what
Hilbert called `proper energy theorems'. Physically in such
theories one has a localized, conserved energy density; and 
one can prove that in any arbitrary volume the net outflow of energy 
across the boundary is
equal to the time rate of decrease of energy within the volume. As will
be shown below, this follows from the fact that the energy-momentum tensor
of the theory is divergence free. In general relativity,
on the other hand, it has no meaning to speak of a definite localization of
energy.   One may define a quantity which is divergence free analogous to the energy-momentum density tensor of special relativity , but it  is gauge dependent: i.e., it is not covariant under general coordinate transformations. 
Consequently the fact that it
is divergence free does not yield a meaningful law of local energy conservation.
Thus one has, as Hilbert saw it, in such theories `improper
energy theorems'.

A key feature for physics of Noether's {\it{I.V.}} paper is the clarity 
 her theorems brought to our understanding of the principle of
energy conservation. As Feza Gursey wrote \cite{feza}: \lq\lq Before Noether's Theorem the principle of conservation of energy was shrouded in
      mystery, leading to the obscure physical systems of Mach and Ostwald. Noether's simple and
      profound mathematical formulation did much to demystify physics."
Noether showed in her theorem I that the principle of energy conservation follows
from symmetry under time translations. This applies to theories
having a finite continuous symmetry group; theories that are Galilean or
Poincar\`{e} invariant, for example.
In general relativity,
on the other hand, 
energy conservation takes a different form as will be shown
below. Noether's theorem II applies in the case of general relativity
and one sees that she has proved Hilbert's assertion that in this case one
has  `improper energy theorems', and that this is a characteristic
feature of the theory.  It is owing to the fact that the  
theory is a gauge theory; i.e., that it has an infinite continuous
group of symmetries of which time translations are a subgroup. Indeed 
generally she defines as \lq\lq improper"
 divergence relationships, which vanish when the field equations are satisfied, 
which  correspond to a finite continuous subgroup of an infinite continuous group.
Generally they
do not have the required invariance or covariance properties under the
larger group. For example,
in general relativity a divergence
free energy-momentum (pseudo) tensor can be constructed but it is
gauge dependent (see below). Because  it is not covariant under general coordinate
transformations, it is more properly called  a pseudotensor. 
Such pseudotensors are covariant with respect to the linear transformations of the Poincar\`{e} group and may be used
in  asymptotic spacetime regions far from gravitating sources to derive a principle of
energy conservation.\cite{c&w}

To illucidate these matters further, we discuss in some detail 
  field theories of matter, gravity, electromagnetism, etc. in both
special and general relativity.  In special relativity these theories have a
`proper energy theorem' in the sense of Hilbert and we will show how `proper energy theorems' give a
principle of
local energy conservation.   In general relativity, on the other hand,
the proper energy theorem becomes improper in that the energy-momentum
tensor for which the theorem holds is gauge dependent. As will be shown
below,
there is transfer
of energy to and from the gravitational field and it has no meaning
to speak of a definite localization
of the energy of the gravitational field in space.\footnote{A 
clear exposition regarding gravitational field energy is given by Landau and Lifshitz\cite{L&L}.} Consequently we do not have a principle of local
energy conservation in spacetime regions in which there exist gravitational
fields.

The theories we discuss here  are field theories which may be formulated in
terms of a variational principle; i.e., their field 
 equations may be obtained from Hamilton's principle.
This was Hilbert's great contribution to the general theory of relativity. He showed the correct field equations can be derived from  Hamilton's 
principle, and   Noether's work followed upon this. 
In this formulation, the theory is defined by a Lagrangian $\cal{L}$. The
field equations are derived from Hamilton's principle that the action
\begin{equation} 
I = \int dx ~\cal{L} \label{act}
\end{equation}
is an extremum. The Lagrangian depends on the field quantities and their
derivatives. Symbolically let $\phi$ represent the field quantities and,
for simplicity,
 assume $\cal{L}$ depends only on $\phi$ and its first derivatives.
Then Hamilton's Principle gives 
 the field equations: viz.,  
 \begin{equation} {\delta {\cal{L}}
\over \delta \phi} - {\partial \over \partial x^\nu}{\delta{\cal{L}} \over
\delta\partial \phi/\partial x^\nu } ~ = ~ 0 \; .  \label{symFE} 
\end{equation}
Here we discuss in some detail laws of energy
conservation in special and general relativity where Noether's theorem I
applies in the first case and theorem II in the second. \footnote{
 In the next section, Noether's
theorems  will be described more generally.}
 When the action is invariant with respect to 
transformations of the Poincar\'{e} 
group,   Noether's theorem I gives a divergence-free
 energy-momentum  tensor $T^{\mu\nu}(x)$ which, symbolically in the notation of 
  (\ref{symFE}), for scalar fields is given by 
\footnote{ We use Minkowski notation for the components
of the spacetime point $x$; viz., $x^\mu$ with $\mu = 0, 1, 2, 3$ with $x^0 = t$
and $x^i$ for three orthogonal space directions ($ i = 1, 2, 3$)~;
 $g^{\mu\nu}$ is the Minkowski metric tensor whose components are 
$(-1, 1, 1, 1)$. We also use the notation $\partial_{\mu} \equiv \partial / 
\partial x^\mu $~.} 

\begin{equation}
T^{\mu\nu} = {\delta {\cal{L}}\over \delta \partial_{\mu} \phi} \partial^\nu 
\phi + g^{\mu\nu} {\cal{L}}\; .  \label{Tmunu} 
\end{equation}
More generally one may write the energy-momentum tensor as
\footnote{This form is useful  later when we generalize to general relativity.}

\begin{equation} 
 T^{\mu\nu} = - 2 {\delta{\cal{L}} \over \delta g_{\mu\nu}} ~ 
 +~g^{\mu\nu}{\cal{L}} \; .
  \label{tmunu}
  \end{equation}
 Theorem I shows 
 invariance of the action  under spacetime translations implies
 that the divergence of  $T^{\mu\nu}$ is 
 proportional to the left  side  of (\ref{symFE})\footnote{This is called by
 Noether a Lagrange function.}. Consequently
 when the field equations are satified, the divergence vanishes and we have 
 \footnote{Repeated
indices are summed. Relativists' notation for this divergence is 
$T^{\mu\nu} ~_{,\nu} = 0.$}
\begin{equation}
\partial_\nu T^{\mu\nu} = 0 \; .   \label{vd}
 \end{equation}
  This  implies local energy 
conservation.
    The energy
density at the spacetime point $x$  is given by the
   $T^{00}(x)$  
 and the $T^{i0}$ components give the energy crossing a unit area per
unit time in the direction of the normal $i$. The time-like equation (\ref{vd}),
 i.e., $ \partial_\nu T^{0\nu} = 0$, follows from the invariance of the
theory under time translations.
To show that   $\partial_\nu T^{0\nu} = 0$ gives local energy conservation,  it is convenient to
use the 3-vector $\vec{\cal{F}}$ for the $T^{0i}$ components; then
(\ref{vd}) reads 
\begin{equation} 
- ~\partial T^{00}/\partial t = div~ \vec{\cal F} \;.
\label{divf} 
\end{equation} 
Integration of (\ref{divf}) over any 
spatial volume $V$ shows, by Gauss's theorem, that the rate of change of
energy in $V$ equals the net energy flow across the bounding surface
 per unit time.
This holds for arbitrary and, therefore, arbitrarily small volumes. Thus
(\ref{vd}) gives, in the language of 
contemporary
physicists,   local energy conservation.  This is a
`proper energy theorem' in the sense of Hilbert.

In general relativity, on the other hand, the components of the metric tensor $g^{\mu\nu}$
 are the gravitational field variables and generally  spacetime dependent.
They are included as field variables in  
the Lagrangian of the theory with the addition of the Hilbert-Einstein term $R$;
  ${\cal{L}}$  has in it, as before, the electromagnetic, matter, etc. field variables.
The action is 
\begin{equation}
I =  \int d^4x \sqrt{-g} ~( {\cal{L}} + \chi^{-1}R )  \label{gravac}
\end{equation}
where $g$ is the determinant of $g^{\mu\nu}$, $\chi$ is a
coupling constant and 
$R$ is the Riemannian curvature scalar.
Following Hilbert, one may derive  the field equations of
general relativity from the condition that  (\ref{gravac}) is
 an extremum
with respect to variations of  $g^{\mu\nu}$. One obtains 
\begin{equation} 
R^{\mu\nu} - \case 1/2Rg^{\mu\nu} = \chi ~
T^{\mu\nu} \label{FE} 
\end{equation}
where $R^{\mu\nu}$ is the Ricci tensor constructed by contraction from 
the Riemann curvature tensor,
and $R \equiv g_{\mu\nu}R^{\mu\nu} $ is the Ricci or curvature scalar.  
 The Riemann
curvature tensor is constructed from the metric tensor $g^{\mu\nu}$ and its first and second
derivatives.  The source term
on the right  side of (\ref{FE})is the  
 energy-momentum tensor $T^{\mu\nu}$ given by  (\ref{tmunu}).
\cite{c&w}~ It contains contributions from the
electromagnetic fields, matter fields, etc. but not from gravitational fields.
Since gravitational fields carry energy and momentum,
 $T^{00}$ is not  the total energy density. As will be shown below,
 $T^{\mu\nu}$ is not divergence-free. 
 It cannot and does not play the same role as regards
energy-momentum conservation as  in special relativity.   This surely was 
immediately clear to Hilbert in 1915. What was apparently not clear 
to Hilbert and Klein in the early days was
how the principle of conservation of energy is satisfied in this theory. This will be discussed below.

For theories whose symmetry group is an infinite continuous group, the main
results of theorem II are that there are certain identities, or `dependencies' as she called them
\cite{noet}, between 
Lagrange functions of the
theory and their derivatives.  This is quite different from the results of
theorem I that for each infinitesimal generator of the finite
continuous group there is a quantity whose divergence vanishes when the
 the Lagrange functions vanish (field equations are satisfied).
In general relativity, the `dependencies'  give the Bianchi
identities for components of the curvature tensor. These results
are covariant under the full group.
In particular, one has 
the contracted Bianchi identity
\begin{equation}
(R^{\mu\nu}  - \case 1/2Rg^{\mu\nu})_{;\mu} ~=~0    \label{bian}
\end{equation}
This equation states that the covariant divergence of the left side of (\ref{FE}) vanishes.
The covariant divergence differs substantially from the ordinary divergence
which we have been discussing up to now.  For a second rank tensor, e.g. 
$T^{\mu\nu}$,
the covariant divergence
  is given by
 \begin{equation}
T^{\mu\nu} ~  _{ ;\nu} = \Gamma^\mu_{\sigma\nu} T^{\sigma\nu} +(\sqrt{-g}
T^{\mu\nu}) ~ _{,\nu}/\sqrt{-g} \label{covd}
\end{equation}
 where $\Gamma^\mu_{\sigma\nu}$ are Christoffel symbols which depend upon
 $g^{\mu\nu}$ and its derivatives and are
non-vanishing when the Riemann curvature is different from zero.
The covariant divergence is an invariant; i.e., the covariant divergence
of a second rank tensor gives a first rank tensor. Thus (\ref{bian})
 implies
\begin{equation}
T^{\mu\nu} ~  _{ ;\nu} = ~0 \; \label{covcons}
\end{equation}
This is often referred to as the  energy-momentum conservation law
for general relativity. The presence of the metric tensor and its
derivatives in (\ref{covd}) shows the energy-momentum transfer 
contained in (\ref{covcons}) between the gravitational
fields and the source fields in $T^{\mu\nu}$.

From
(\ref{covd}) it is clear the the ordinary divergence $T^{\mu\nu}~_{,\mu}$
is in general different from zero in spacetime regions with gravitational
 fields. It is interesting to note, however, that 
at any given spacetime point one may choose a set of coordinates for which
the gravitational fields vanish  ($g^{\mu\nu}$ reduces to the flat spacetime
Minkowski metric and the Christoffel symbols vanish). This is
guaranteed by the equivalence
principle which states that one can always choose a coordinate system such 
that spacetime in the neighborhood of a given
point is Minkowski (flat). Thus one may see why it is not meaningful to speak
of a localized energy density for gravitational fields.

The physics of these relations is somewhat complicated. In regions of spacetime
near gravitating sources, where the Riemann curvature is non-vanishing,
there is  failure of a principle of local energy conservation. The energy balance locally
cannot be discussed independently of the coordinates one uses to calculate
it, and consequently different results are obtained in various different
coordinate frames - some being artifacts of the calculation itself.
An amusing example of this is the case of an accelerating
mass.  For simplicity, let us take the mass and acceleration small, so that the
deviations from Newtonian physics are small and may be treated perturbatively -
the so-called post-Newtonian approximation.
In this approximation the theory is  similar to electromagnetic theory.  In
electromagnetic theory, an accelerating charge radiates and there is the radiation
reaction force which tends to deccelerate it; there is energy loss to the
radiation field.  In Einstein gravity, the accelerating mass is a source of
gravitational radiation and one may calculate the radiation reaction force as
one does in electromagnetic theory.  A naive calculation along these lines
yields a radiation reaction force term  (\.{a}
term in the equation of motion with a the acceleration)that has the opposite sign to the electromagnetic
case; the particle appears to speed up gaining energy as it
radiates!  This is an unphysical radiation reaction force.
The calculation is not gauge invariant and  depends on the
choice of coordinates used to describe the motion of the particle.

However, though local energy conservation fails, there is a large scale
principle of
energy-momentum conservation in the general theory.  
The
 problem in principle was solved by Noether.  As she pointed out in {\it{I.V.}}, it is  possible to construct a
divergence-free 
quantity generally referred to as the 
 a pseudotensor  $T^{\mu\nu}_{eff}$.  Jackiw et al. \cite{jackiw}
have recently constructed such a pseudotensor using her theorem II.
 Other constructions have been made by 
Weinberg \cite{wein} and Landau and Lifshitz \cite{L&L}. 
Such pseudotensors have the form 
\begin{equation}
T^{\mu\nu}_{eff} = T^{\mu\nu} + t^{\mu\nu} \; .  \label{teff}
\end{equation}
and satisfy 
 \begin{equation}
 T^{\mu\nu}_{eff} ~ _{,\mu} = 0 \label{divfeff}
\end{equation}
The $ t^{\mu\nu}$ are not unique, and they are gauge dependent.
Owing to this gauge dependence, $T^{\mu\nu}_{eff}$  
 is called a pseudotensor because it does not transform as a second rank tensor
under the full group of general coordinate transformations.
However, these constructions are covariant under the linear transformations
of the Poincar\'{e} group and may be used to calculate energy and momentum 
 in gravitationally field free regions and asymptotically field free regions.
Using such the pseudotensor
$T^{\mu\nu}_{eff}$,  one can formulate a principle of energy conservation by integrating over a large,
asymptotically flat spacelike hypersurface.
\footnote{ This is done in Refs. \cite{c&w} and (\cite{L&L}.}  In so doing one obtains a total energy-momentum
four-vector $P^\alpha$ which is constant in time and includes  the
energy and momentum carried by gravitational radiation.  The $t^{\mu\nu}$
contributions take gravitational radiation into account, and the calculations
are done over distances large compared to the wavelength of the radiation.  This
total energy-momentum vector behaves as a special relativisic four-vector for
any transformation which is a Lorentz transformation far from sources and
 the vector is independent of the choice of coordinates used
in the calculation.\cite{c&w} Such a calculation is applied to the observed
decrease of the orbital period of the binary pulsar PSR 1913+1916, and used to
estimate the associated flux of gravitational radiation.\footnote{ A detailed
discussion of this and references to the literature can be found in Ref.
\cite{c&w}.}  
\section{The Theorems}

The main results of {\it{I.V.}} are two theorems which are proved along
with their converses.
  They
will be described here. Felix Klein's statements \cite{basil} of the theorems
 are given  in Appendix B. 

Though they apply more generally, we shall describe the
theorems as they apply to field theories. Consider an integral
${\bf{I}}$
over independent variables $x$ of a functional $\cal{L}$ depending on
functions (field variables) of these $x$ variables $\phi(x)$ and  their derivatives; viz.,
\begin{equation}
{\bf{I}} = \int dx {\cal{L}}( \phi, \partial \phi/ \partial x, \partial^2 \phi/
\partial x ^2, ....) \; .                                                          
\end{equation}
 The
functional $\cal{L}$ may also depend directly on the $x$.
 The integral
 {\bf{I}} in physics 
is the action in the sense of Hamilton's Principle, 
and the functional  $\cal{L}$ is the Langrangian  of the theory.
The theorems apply to systems for which ${\bf{I}}$ is an invariant 
of a Lie group of transformations which transform $x \rightarrow y,
\phi \rightarrow \chi, ... $ ~ ; i.e., when one has
\begin{eqnarray}
{\bf{I}} =  \int dx {\cal{L}}(x, \phi, \partial \phi/ \partial x, \partial^2 \phi/
\partial x ^2, ....)   \nonumber                  \\
& & \\
& = & \int dy {\cal{L}}(y, \chi, \partial \chi/ \partial y, \partial^2 \chi/
\partial y^2, ....) \nonumber
\end{eqnarray}
integrated over an arbitrary region of $x$-space and the corresponding
$y$-space.

The two cases of a finite and an infinite Lie group are separately considered.
In Noether's notation,   a finite Lie group is denoted by $\cal{G}_\rho$; it is a continuous group  of transformations
whose most general transformation depends analytically on a 
finite (or countably infinite) number $\rho$ of independent 
parameters. An infinite Lie group, on the other hand, is denoted by 
$\cal{G}_{\infty\rho}$. It 
is a continuous group 
whose most general transformation depends on $\rho$ independent, arbitrary
 {\it{functions}} and their derivatives.  The results of theorem I  apply when the action ${\bf{I}}$
is an invariant 
 of a finite Lie group, and the results of theorem II when it is an invariant of
an infinite Lie group. The results are different in the two cases.
                                                    
Theorem I for $\cal{G}_\rho$ gives the result that there is a  divergence
 for each of the $\rho$ 
independent generators of the  group which is 
proportional to a Lagrange function. In her terminology the Lagrange functions
of the theory are the left hand side of equations of motion (field equations)
of which
the right hand side  vanishes. Thus when the equations of motion 
are satisfied, the divergences vanish. For the spacetime displacements of the
Poincar\'{e} group, the four divergences are four  tensor component 
divergences. The vanishing of these divergences give energy-momentum conservation.
The U(1) Lie group of global gauge transformations 
in   electrodynamics  is another example.  Corresponding to the generator of this group, one has
a divergence free  current and from this one proves charge conservation.
\footnote{If, on the other hand, the action is invariant with respect to
 local gauge transformations, the symmetry is of an infinite Lie group 
corresponding to the possibility of independently transforming the gauge
at every spacetime point. Theorem II applies in this case. As in theorem I
case, one  gets
electric current conservation from theorem II. However, if non-Abelian
gauge transformations are considered, the result is more complicated.}

Theorem II applies when the action is an invariant of an infinite Lie group 
$\cal{G}_{\infty\rho}$. The statement of the theorem for this case is very general and, aside from its application
to general relativity, it applies in a wide variety
of other cases.   For example,  quantum chromodynamics and other gauge field theories are theories to which it applies.  
From theorem II, one has
identities  between Lagrange functions and their derivatives. 
See Appendix B.
These identities 
Noether calls `dependencies'. Generally they are covariant with respect to the
group. For example,
 the Bianchi identities in the general theory of relativity are 
examples of such `dependencies'.

It is important to point out that in {\it{I.V.}} proofs are given for the converses of  theorems I and II as well.
 For example, theorem I states that if the action
${\bf{I}}$
is an invariant of a finite Lie group $\cal{G}_\rho$ , there are $\rho$
divergences which vanish when the equations of motion are satisfied.
The converse  also is proved in {\it{I.V.}}; namely that
 the $\rho$ vanishing divergences imply that the action is an invariant of 
$\cal{G}_\rho$. This result was important in the history of discovery
of symmetries in elementary particle physics.\cite{byers}

I conclude this talk with a quotation from {\it{I.V.}} where
    Noether says that  Theorem II is  \lq\lq  proof
of a Hilbertian assertion about the connection of the failure of
proper laws of conservation of energy with general relativity ... in
a generalized
group theory version."  One of the reasons why her theorems have been very important for 20th century  physics is because they are in `a generalized group theory version'.

\acknowledgements
I would like to thank Mina Teicher and her colleagues for organizing this very interesting and rewarding Symposium; and my colleague Basil Gordon for the help he afforded me in understanding and appreciating the workings of the minds of these
great mathematicians, and for his English translations of their papers,
and for his close reading of this manuscript in its various drafts. 
I would also like to thank my colleague Aaron Grant for a critical reading of
the final draft.
\newpage
\appendix{APPENDIX A}

\begin{enumerate}
\item{} \"Uber die Bildung des Formensystems der tern\"aren 
biquadratischen Form.
  Sitz. Ber. d. Physikal.-mediz. Soziet\"at in Erlangen 39
 (1907), pp. 176-179.
\item{}  \"Uber die Bildung des Formensystems der tern\"aren 
biquadratischen Form.
   Journal f. d. reine u. angew. Math. 134 (1908),
 pp. 23-90.
\item{} Zur Invariantentheorie der Formen von n Variabeln.
  J. Ber. d. DMV 19 (1910), pp. 101-104.
\item{} Zur Invariantentheorie der Formen von n Variabeln.
  Journal f. d. reine u. Angew. Math. 139 (1911), pp. 118-154.
\item{} Rationale Funktionenk\"orper.
J. Ber. d. DMV 22 (1913). pp. 316-319.
\item{} K\"orper und Systeme rationaler Funktionen.
  Math. Ann. 76 (1915), pp. 161-196.
\item{} Der Endlichkeitssatz der Invarianten endlicher Gruppen.
  Math. Ann. 77 (1916), pp. 89-92.
\item{} \"Uber ganze rationale Darstellung der Invarianten
 eines Systems von beliebig vielen Grundformen.
  Math. Ann. 77 (1916), pp. 93-102. 
 \item{} Die allgemeinsten Bereiche aus ganzen 
 transzendenten Zahlen.
  Math. Ann. 77 (1916), pp. 103-128. 
 \item{} Die Funktionalgleichungen der isomorphen Abbildung.
  Math. Ann. 77 (1916), pp. 536-545.
 \item{} Gleichungen mit vorgeschriebener Gruppe.
  Math. Ann 78 (1918), pp. 221-229. 

\end{enumerate}
\newpage

\appendix{APPENDIX B}

\noindent
The following quotation is from Felix Klein's annotation to his paper \lq\lq{\it{\"{U}ber die Differentialgesetze f\"{u}r
die Erhaltung von Impuls und Energie in die Einsteinschen
Gravitationstheories}}" \cite{kleinpap} which can be found in his
annotataed collected works 
 {\it{Gesammelte mathematische Abhandlungen}}, Julius Springer, Berlin 1921;
Ch. XXXI (erster band, p. 559).  In this quotation 
  \lq\lq The main theorem given in section 2 above" refers to his  paper which he is annotating.

\lq\lq I presented the main theorems of Frl.  Noether at the July 26 meeting ...
The main theorem given in section 2 above is a special case of the following
far-reaching theorem of Frl. Noether:
\begin{quote}
{\it{ If an integral I
is invariant under a continuous group $G_\rho$ with $\rho$ parameters, then
$\rho$ linearly independent combinations of the Lagrangian expressions are
divergences.}}   
\end{quote}
Concerning in particular the claim made by Hilbert
[that the failure of `proper' energy theorems was  characteristic
  of the theory], the following is an exact formulation
according to Frl.  Noether:  
\begin{quote}{\it{ If an integral $I$ is invariant under
the translation group, the energy relations are improper if and only if $I$ is
invariant under an infinite-dimensional group which contains the translation
group as a subgroup.  }} 
\end{quote}
Incidentally, the theorem of Hilbert that
there are four relations between the field equations 
[Bianchi identities]
also finds its
generalization in Frl.  Noether's work.  Her theorem is as follows:
{\it{ If an integral $I$ is invariant under a group with $\rho$
arbitrary functions in it with their derivatives up to the $\sigma$'th order,
then there are $\rho$ identities between the Lagrangian expressions and their
derivatives of order up to $\sigma$.}}" 

\noindent
English translation courtesy of Basil Gordon.

\onecolumn


\baselineskip=20pt
\begin{references}


\bibitem{noet} E. Noether, 
\lq\lq{\it{ Invariante Variationsprobleme}}", Nachr. d. K\"{o}nig. Gesellsch. d.
Wiss. zu G\"{o}ttingen, Math-phys. Klasse (1918), 235-257; English translation
M. A. Travel, Transport Theory and Statistical Physics 1(3) 1971,183-207.

\bibitem{byers} Nina Byers, {\it{History of Original Ideas and Basics Discoveries
in Particle Physics}}, H. B. Newman and T. Ypsilantis ed., Plenum Press, New York (1996).

\bibitem{klein} Correspondence between D. Hilbert and F. Klein published by
F. Klein in 
Nachr.  d. K\"{o}nig. Gesellsch. d.
Wiss. zu G\"{o}ttingen, Math.-phys. Klasse (1918) as 
{\it{Zu Hilberts erster Note \"uber die Grundlagen der Physik}};  included
  in his annotated collected works Ref. \cite{felix}; see
Ch. XXXI (erster band, p. 560).  English translation courtesy of Basil Gordon.

\bibitem{hilbert1} D.Hilbert, {\it{Gesammelte Abhandlungen}}, Springer-Verlag, New
York, 1970.
 
\bibitem{publist} {\it{Emmy Noether, Collected Papers}}, Springer-Verlag 1983.


\bibitem{paper1} E. Noether, \lq\lq{\it{ Invarianten beliebiger
Differentialausdr{\"{u}}cke,}}", Nachr. d. K\"{o}nig. Gesellsch. d.
Wiss. zu G\"{o}ttingen, Math-phys. Klasse (1918) 37 - 44.

\bibitem{weyl} H. Weyl,
 {\it{Scripta Mathematica III.}} 3 (1935) 201-220; an English translation of
this memorial lecture is given 
 in Ref. ~\cite{Dick}.

\bibitem{Dick} Auguste Dick,{\it{ Emmy Noether (1882 -1935)}},  Birkhauser 1981; English translation by  H. I. Blocher.


\bibitem{pais} A. Pais, {\it{Subtle is the Lord}}, Oxford University Press 
1982.


\bibitem{einstein} A. Einstein, {\it{Sitzungsberichte,}} Preussische Akademie
der Wissenschaften, 1915, p. 844.


\bibitem{hilbert} D. Hilbert, \lq\lq{\it{Die Grundlagen der Physik}}," 
G\"ottinger Nachrichten, Math.-phys. Klasse (1915), S. 395-407.



\bibitem{kimb} Clark Kimberling in {\it{Emmy Noether, A Tribute to Her Life 
and Work}}; James W. Brewer and Martha K. Smith, ed.; Marcel Dekker, Inc.



\bibitem{reid} Constance Reid, {\it{Hilbert}}, Springer-Verlag, New York
(1970).


    

\bibitem{kleinpap}  F. Klein, "{\it{\"{U}ber die Differentialgesetze f\"{u}r
die Erhaltung von Impuls und Energie in die Einsteinschen
Gravitationstheories.}}", Nachr. d. K\"{o}nig. Gesellsch. d.
Wiss. zu G\"{o}ttingen, Math-phys. Klasse (1918).



\bibitem{felix} Felix Klein, {\it{Gesammelte Mathematische Abhandlungen}},
Julius Springer, Berlin 1921.

\bibitem{basil}
English translation 
courtesy of Basil Gordon.


\bibitem{schm}E. Noether and W. Schmeidler," {\it{Moduln in nichtkommutativen Bereichen,
 insbesondere aus Differential-
  und Differen-zenaus-dr\"ucken}}", Math.
 Zs. 8 (1920), pp. 1-35.

\bibitem{feza} Feza Gursey as quoted by Nathan Jacobson in his lengthy Introduction to {\it{Emmy Noether, Collected Works}},
Ref. \cite{publist}.

\bibitem{c&w} See, e.g., Ignazio Ciufolini and John Archibald Wheeler,{\it{
Gravitation and Inertia}}, Princeton University Press, Princeton, N.J.
1995.

 
\bibitem{L&L} L. Landau and E. Lifshitz, {\it{The Classical Theory of Fields}},
English translation by M. Hamermesh, Addison-Wesley Publishing Co., Inc.,
Reading, MA 1951. 


\bibitem{jackiw} D. Bak, D. Cangemi, and R. Jackiw, {\it{Phys. Rev.}}D 49: 5173
(1994).

\bibitem{wein} Steven Weinberg, {\it{Gravitation and Cosmology}}, John Wiley \& 
Sons, Inc., New York 1971.

\end{references}
\end{document}